# Oxygen vacancies: the origin of n-type conductivity in ZnO

Lishu Liu[1], Zengxia Mei[1]*, Aihua Tang[1], Alexander Azarov[2], Andrej Kuznetsov[2], Qi-Kun Xue[3]* and Xiaolong Du[1]*

[1]Key Laboratory for Renewable Energy, Beijing National Laboratory for Condensed Matter Physics, Institute of Physics, Chinese Academy of Sciences, Beijing 100190, China. [2]Department of Physics, University of Oslo, Oslo P.O. Box 1048, NO-0316, Norway. [3]Department of Physics, Tsinghua University, Beijing 100084, China
*e-mail: zxmei@iphy.ac.cn; qkxue@mail.tsinghua.edu.cn; xldu@iphy.ac.cn

**Oxygen vacancy ($V_O$) is a common native point defects that plays crucial roles in determining the physical and chemical properties of metal oxides such as ZnO. However, fundamental understanding of $V_O$ is still very sparse. Specifically, whether $V_O$ is mainly responsible for the n-type conductivity in ZnO has been still unsettled in the past fifty years. Here we report on a study of oxygen self-diffusion by conceiving and growing oxygen-isotope ZnO heterostructures with delicately-controlled chemical potential and Fermi level. The diffusion process is found to be predominantly mediated by $V_O$. We further demonstrate that, in contrast to the general belief of their neutral attribute, the oxygen vacancies in ZnO are actually +2 charged and thus responsible for the unintentional n-type conductivity as well as the non-stoichiometry of ZnO. The methodology can be extended to study oxygen-related point defects and their energetics in other technologically important oxide materials.**

Among metal oxides, ZnO is a prototypical n-type material with numerous applications, including catalysts, gas sensors, varistors and transparent electrodes etc. Renewed interest has recently emerged for its ultraviolet light emission capabilities[1]. Progress towards ZnO-based optoelectronic devices and applications, however, has been impeded largely by unintentional and seemingly unavoidable n-type conductivity of ZnO materials that makes stable p-type doping extremely daunting. In order to solve this most puzzling problem for ZnO, it is highly desirable to understand the origin of the unintentional n-type conductivity[2–6].

It is generally accepted that the oxygen vacancy ($V_O$) plays a central role in determining the physical and chemical properties of metal oxides: it dominates various diffusion mechanisms involved in doping and its limitation, processing and device degradation, minority carrier lifetime, and luminescence efficiency etc[7–13]. Specifically, $V_O$-related issues are of particular interest for ZnO. Whether the above-mentioned n-type conductivity comes from $V_O$ has remained as a major problem in ZnO for a long time: O-deficient ZnO easily behaves as an n-type semiconductor even without the introduction of any intentional dopants[14]. Theoretical studies[15–17] seem consistently to indicate that $V_O$ has the lowest formation enthalpy among the donor-like point defects. However, it leads to a deep electronic state in ZnO so that $V_O$ could not be ionized and contribute to the conductivity at room temperature. The theoretically derived $V_O$ defect energetics exhibits huge controversy probably owing to the finite-supercell formalism and the inaccurate description of the band structure of ZnO[18]: the formation enthalpy $\Delta H_f$ of neutral $V_O$ under O-poor conditions ranges from -0.8 to 3.9 eV[15–17]. A formation enthalpy $\Delta H_f \geq 3.9$ eV as reported in Ref. 15 implies a $V_O$ concentration $C(V_O) \leq 10^8$ cm$^{-3}$, which is far below the experimental value of $10^{17}$ - $10^{19}$ cm$^{-3}$ [19–21]. In the experimental side, most proposed models rely on some circumstantial methods and evidences[14,22]. Moreover, the charge state of $V_O$ is much less addressed and remains as a controversial issue in ZnO.

To resolve these problems, we conduct a systematic study of oxygen self-diffusion in ZnO. Our study employs a unique oxygen-isotope ZnO heterostructures with delicately-controlled chemical potential ($\mu$) and Fermi level ($E_F$). Unlike the gaseous-exchange technique, which has absorption and evaporation problems[23,24], the self-diffusion in the isotope heterostructures provides reliable information on the energetics and electronic properties of point defects[25–27]. We show unambiguously that the diffusion process of oxygen atoms in ZnO is predominantly mediated by $V_O$ rather than oxygen interstitial ($O_i$). We further demonstrate that the oxygen vacancies in ZnO are +2 charged even near the conduction band minimum, in contrast to the general belief of their neutral attribute. The results establish that the oxygen vacancies are the dominant donor-like native point defects and thus responsible for the unintentional n-type conductivity as well as the non-stoichiometry of ZnO.

It is well known that diffusion phenomenon is closely associated with defect formation process. The defect formation enthalpy $\Delta H_f$ is given by

$$\Delta H_f(D^q, \mu, E_F) = E(D^q) - E_p + \mu + q(E_F + E_{VBM}), \quad (1)$$

where $E(D^q)$ is the total energy of the semiconductor with defect D in a charge state of q, and $E_p$ is the energy of the perfect host. $\mu$, $E_F$ and $E_{VBM}$ are the atomic chemical potential, the Fermi level and the valence band maximum, respectively. Thus, the value of q, i.e., the charge state of a specific defect $D^q$, can be deduced from the dependence of $\Delta H_f$ on $E_F$.

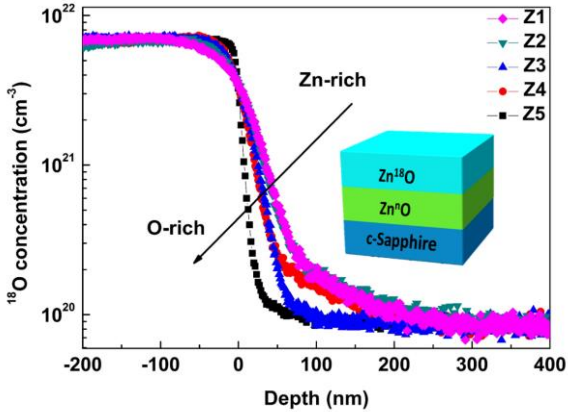

**Figure 1 | $^{18}$O concentration depth profiles in as-grown samples.** Samples labeled as Z1 to Z5 were grown with decreased Zn/O ratio. The inset shows the schematic structure of these samples.

To confirm the role of $V_O$ in the n-type conductivity, the first thing is to determine whether and to what extent it is involved in the diffusion in the heterostructures. As can be seen from Eq. (1), $\Delta H_f$ is linearly related to $\mu$[28], and will increase and decrease in $V_O$- and $O_i$-dominated diffusion processes, respectively, when $\mu$ moves towards O-rich conditions and the influence of $E_F$ and other factors is negligible. A series of heterostructures consisting of top $^{18}$O-enriched $Zn^{18}O$ and bottom natural $Zn^nO$ layers on sapphire were thus synthesized with different chemical potentials and same Fermi levels, labeled as Z1, Z2, Z3, Z4 and Z5, respectively, as schematically illustrated in the inset of Fig. 1(a). The modulation of chemical potentials was realized through a gradually decreased Zn flux under a constant O flux. According to the growth rate, Z1 and Z2 can be regarded as Zn-rich samples while Z3 – Z5 O-rich ones[29] (Supplementary Information Fig. S1). The curves in Fig. 1 show the $^{18}$O concentration depth profiles for the as-grown samples. The enriched $Zn^{18}O$ layers are immediately evident by the higher concentration of $^{18}$O with respect to that in the bottom $Zn^nO$ layers. Interestingly, the $^{18}$O profiles obviously become steeper when the $\mu$ moves towards O-rich conditions.

According to the Fick's law, the self-diffusion profile of $^{18}$O across an interface is described by[27]

$$C(x) = \frac{C_1 + C_2}{2} - \frac{C_1 - C_2}{2} erf(\frac{x}{2\sqrt{Dt + k}}), \quad (2)$$

where $x = 0$ is at the interface, $C_1$ and $C_2$ are the initial $^{18}$O concentrations at the top and bottom sides of heterostructures, respectively, and erf(x) is the error function. The characteristic diffusion length is $l$ ($l = 2\sqrt{Dt}$), where D is the self-diffusion coefficient (i.e., diffusivity) and $t$ is the annealing time. A correction factor $k$ is introduced in the fitting procedure owing to the initial distribution. The O self-diffusion coefficient D can be therefore derived based on Eq. (2). Figure 2(a) shows the typical $^{18}$O concentration depth profiles of Z5 and their fitting curves as a function of annealing temperature using Eq. (2). The diffusion of $^{18}$O atoms becomes more pronounced at elevated annealing temperature. As illustrated in Fig. 2(b), the samples with gradually varied chemical potentials show different profiles after annealing at 1073 K for 30 min. Note that D has a negatively exponential relation to the activation enthalpy $\Delta H_a$, i.e., $D = D_o \exp(-\Delta H_a/k_B T)$, where $D_0$, $k_B$ and T denote the pre-exponential factor, Boltzmann constant and temperature, respectively. Thus, using the data of D at different temperatures, the activation enthalpies $\Delta H_a$ can be obtained. Arrhenius plots of all of the extracted $D$ values versus the corresponding reciprocal absolute temperatures are presented in Fig. 3. As shown in Fig. 3, $\Delta H_a$ is $1.10 \pm 0.04$ eV, $1.32 \pm 0.05$ eV, $1.53 \pm 0.04$ eV, $1.69 \pm 0.03$ eV and $2.56 \pm 0.06$ eV for Z1-Z5, respectively, when $\mu$ moves towards O-rich conditions.

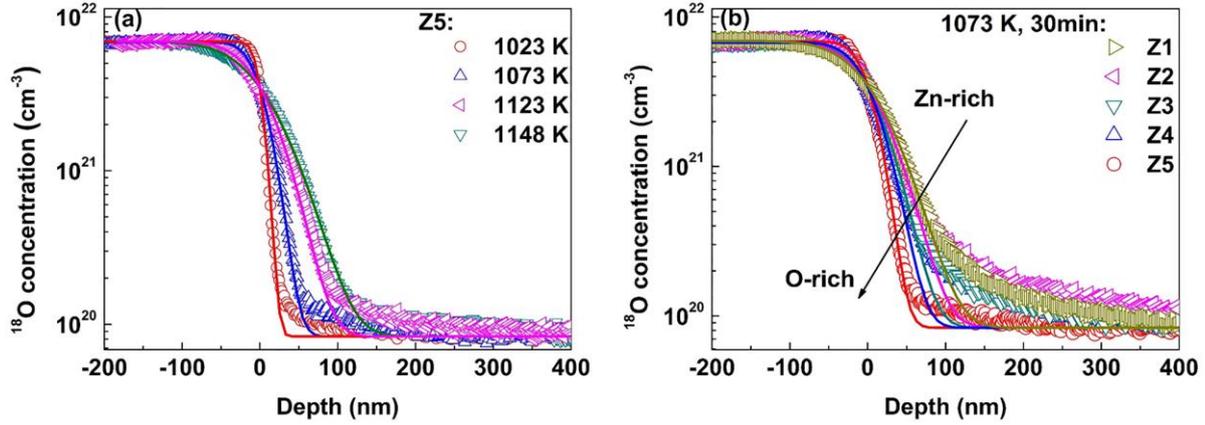

**Figure 2 | Typical concentration depth profiles of $^{18}$O after annealing in vacuum for 30min. (a)** Concentration depth profiles of $^{18}$O with different annealing temperatures ranging from 1023 K to 1148 K for Z5. The solid lines show the fitting results based on Eq. (2). **(b)** Concentration depth profiles of $^{18}$O for Z1-Z5 after annealing at 1073 K for 30 min. The solid lines show the fitting results based on Eq. (2).

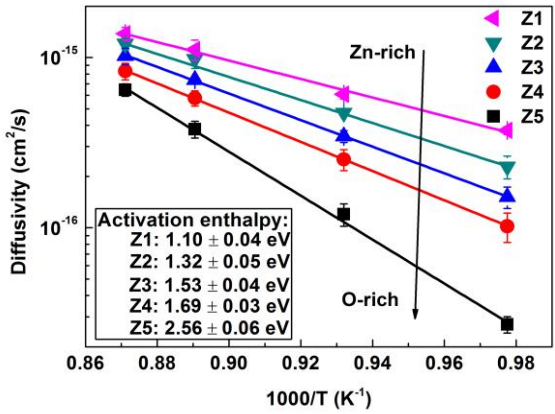

**Figure 3 | Arrhenius plots of the extracted O self-diffusion coefficient $D$ versus the reciprocal temperature 1000/T.** The solid line shows the best fit to the self-diffusion coefficient. The inset shows the obtained activation enthalpy, which increases monotonically from Z1 (Zn-rich) to Z5 (O-rich). Note that the error bars in Fig. 3 can be explained by the inaccuracy originated from the uncertainty in SIMS measurements, given the fact that SIMS sputtering rate is not being constant.

It is well established that $D$ is temperature-dependent and affected by (i) the chemical potential $\mu$ (i.e., the partial pressures of Zn and O), (ii) the Fermi level $E_F$, and (iii) intentional or unintentional doping, i.e., externally controlled mechanism. To determine the dominant diffusion path in Z1-Z5, the influence of $E_F$ and dopants on $D$ has to be excluded first. In this case, $D$ (and hence $\Delta H_a$) will be more predominantly controlled by $\mu$.

In our case, the carrier concentrations of Z1 to Z5 were obtained by temperature-dependent Hall measurements in vacuum from 473 K – 823 K, and reached a nearly identical value, resulting in an almost constant $E_F$ value in the range of 673 K – 823 K (Supplementary information Fig. S2). Note that $E_F$ is estimated based on the well-known formula $E_F = E_F^i + k_B T \ln(n_c/n_i)$, where $E_F^i$, $n_c$ and $n_i$ denote Fermi level under the intrinsic condition, the carrier concentration and the intrinsic carrier concentration, respectively. The temperature dependence of the band gap $E_g$ is also taken into account in the above case[30]. The plots of $E_F$ versus T are then reasonably extrapolated to the annealing temperature region in the diffusion experiments. As a result, an almost fixed $E_F$ is obtained for the five samples Z1 to Z5, suggesting that the influence of $E_F$ on the diffusion coefficient $D$ can be neglected.

Further, the defect concentrations can also be externally controlled, e.g., through intentional or unintentional doping. If impurity concentration is high enough, the dependence of *D* and $\Delta H_a$ on $\mu$ should be significantly less pronounced. However, for our samples, the diffusivity *D* increases by more than one order of magnitude when varying $\mu$ at 1023 K, as illustrated in Fig. 3. Moreover, the concentration of those unintentionally-doped impurities is relatively low (Supplementary information Fig. S3) so that their effect on *D* is negligible. As a result, the diffusivity *D* of the isotopic heterostructures - Z1 to Z5 - is primarily controlled by $\mu$, which is the first goal of our experimental designs. Indeed, $\mu$ can control the defect concentrations and thus influence the diffusion process (Supplementary Information).

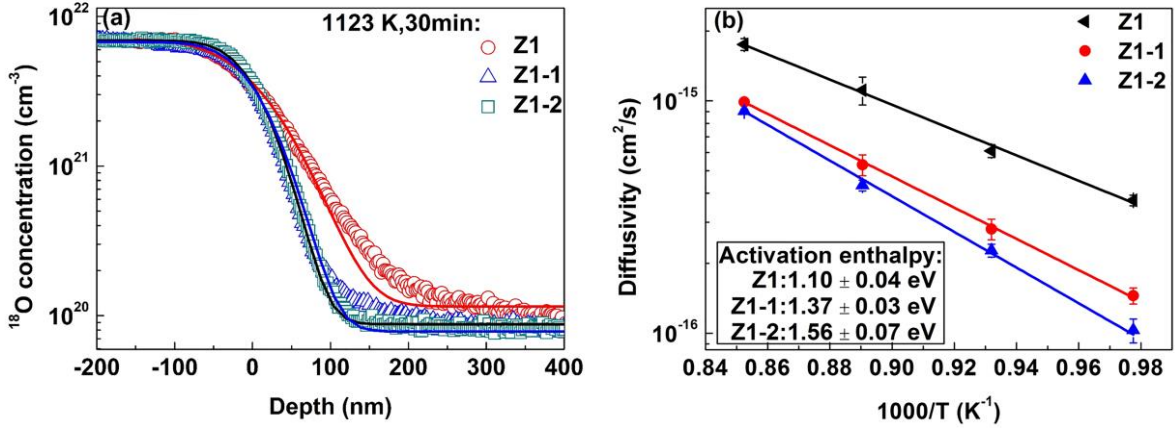

**Figure 4 | Concentration depth profiles and Arrhenius plots of *D* versus 1000/T. (a)** Concentration depth profiles of $^{18}$O for samples with different $E_F$ positions after annealing at 1123 K in vacuum for 30 min. **(b)** Extracted O self-diffusion coefficient *D* versus the reciprocal temperature 1000/T. The inset shows the obtained activation enthalpy, which increases from Z1 to Z1-2. Given the fact that SIMS sputtering rate is not being constant, the error bars in Fig. 4(b) can be explained by the inaccuracy originated from the uncertainty in SIMS measurements.

Under the assumption that the diffusion is a conventional point-defect-mediated process, the activation enthalpy $\Delta H_a$ of the diffusion defect species can be considered as the sum of the formation enthalpy $\Delta H_f$ and the migration enthalpy $\Delta H_m$. The relatively low $\Delta H_a$ (1.10 ± 0.04 eV) for Z1 suggests that a direct diffusion mechanism without the need of forming thermally-activated defects is more reasonable, namely the diffusing atoms simply jump into neighboring vacancy sites. Consequently, the activation enthalpy for Z1 is essentially determined by the migration enthalpy. For Z2 to Z5, on the other hand, the increased $\Delta H_a$ and lower *D* values at identical temperatures shown in Fig. 3 indicate that the defects of mediating diffusion are less favorable under O-rich conditions. In the case that only $V_O$ and $O_i$ are involved in diffusion, $V_O$ should dominate the above-mentioned self-diffusion process. Otherwise, if $O_i$ is dominant, the trend for $\Delta H_a$ and *D* from Z1 to Z5 would be opposite. Moreover, it is consistent with the monovacancy model, in which the migration enthalpy of vacancy $\Delta H_m^V$ is described by[31,32]

$$\Gamma_O = \sqrt{\frac{8\Delta H_m^V}{3m\lambda^2}}, \qquad (3)$$

where $\Gamma_O$, m and $\lambda$ represent the frequency of the highest vibrational mode in the crystal, the mass of O atom and the lattice constant, respectively. Taking $\Gamma_O$ = 8×10$^{12}$ Hz[27,33] and $\lambda$ = 0.52 nm[34] into account, we obtain $\Delta H_m^V$ = 1.08 eV, which is in a good agreement with $\Delta H_a$ = 1.10 ± 0.04 eV for Z1 sample (Fig. 3).

To figure out the dependence of $\Delta H_f$ on $E_F$ and the charge state q of $V_O$, we investigate the influence of $E_F$ on O diffusion. The modulation of $E_F$ was achieved by effective Ga doping. Two more series of samples were studied. The first batch of samples Z1-1 and Z1-2 were grown using the conditions identical to those for Z1, except for *in-situ* Ga doping contents of ~2.0×10$^{18}$ cm$^{-3}$ and ~6.0×10$^{18}$ cm$^{-3}$, respectively (Supplementary information Fig. S4). The second batch of samples Z2-1 and Z2-2 has Ga concentrations of ~$3.0\times10^{18}$ cm$^{-3}$ and ~$1.5\times10^{19}$ cm$^{-3}$, respectively (Supplementary information Fig. S5). As illustrated in Fig. 4(a), the diffusion process is significantly retarded in Ga-doped samples. According to the Arrhenius plots in Fig. 4(b), the activation enthalpy $\Delta H_a$ increases from 1.10 ± 0.04 eV for the undoped sample Z1 to 1.56 ± 0.07 eV for the Ga doped Z1-2. The $\Delta H_a$ for Z2, Z2-1 and Z2-2 shows similar behavior and increases from 1.32 ± 0.05 eV for Z2 to 1.77 ± 0.02 eV for Z2-2 (Supplementary information Fig. S6). After annealing (up to 1123 K) the secondary ion mass spectroscopy (SIMS) measurements do not reveal a significant redistribution of Ga compared to the as-grown samples (Supplementary information Fig. S7), indicating that Ga dopant diffusion is negligible. Since the Zn/O ratio and thus $\mu$ are fixed for each batch of samples, the change in $\Delta H_a$ should be mainly attributed to the Fermi level change after Ga doping. Indeed, the Fermi level is below the conduction band minimum by 0.35 eV, 0.18 eV and 0.12 eV[30] for undoped Z1 and Ga-doped Z1-1 and Z1-2 with carrier concentration of $1.0\times10^{17}$ cm$^{-3}$, $1.0\times10^{18}$ cm$^{-3}$ and $2.3\times10^{18}$ cm$^{-3}$, respectively, at 823 K.

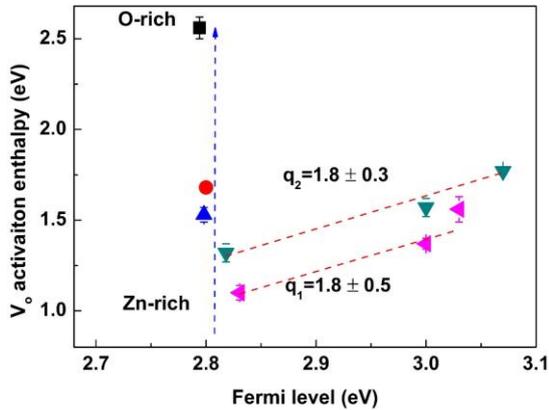

**Figure 5 | Activation enthalpy of V$_O$ as a function of Fermi level.** Dashed lines illustrate an ascending trend for $\Delta H_a$ with increased Fermi level for two batches of samples: Z1 to Z1-2 and Z2 to Z2-2. As shown by the red dashed lines, the charged state of V$_O$ is obtained by fitting the two batches of samples using linear relation, respectively. $\Delta H_a$ as a function of chemical potential is also included for Z1-Z5 samples, as illustrated by the blue dashed lines.

The result of the increased activation enthalpy with the upward Fermi level indicates that V$_O$ is charged, which constitutes another finding of our study. Since $\Delta H_f$ of charged defects is linearly related to $E_F$ and $\Delta H_m$ is independent on $E_F$ and $\mu$ for defects with a given state, $\Delta H_a$ is linearly dependent on $E_F$ for defects with a given state. Accordingly, the slope of $\Delta H_a$ - $E_F$ curve gives the charge number q of the defects. The results for Z1 to Z1-2 and Z2 to Z2-2 are shown in Fig. 5. We can clearly see that $\Delta H_a$ increases with $E_F$, which is in contrast to the theory horizontal lines[15]. The fitting for both batches of samples gives a consistent value of q$_1$, q$_2$ = +1.8, indicating that the dominant V$_O$ is +2 charged even near the conduction band minimum. The ionized V$_O$ is a shallow donor and thus gives rise to the n-type conductivity in ZnO. The obtained $\Delta H_m$ = 1.10 eV therefore corresponds to the migration barrier for ~+2 charged V$_O$, which is excellent agreement with the theoretical result – 1.09 eV[33].

In conclusion, we have investigated the self-diffusion process of oxygen atoms as a function of $\mu$ and $E_F$ in oxygen-isotopic ZnO heterostructures. It is found that V$_O$ dominates the diffusion process with an activation enthalpy of 1.10 eV – 2.56 eV from Zn-rich to O-rich growth conditions. We have deduced the migration enthalpy to be 1.10 eV and found that the formation enthalpy of V$_O$ increases even when $E_F$ moves towards the conduction band minimum. Our results reveal that V$_O$ is +2 charged even near the conduction band minimum, which is oppose to the usually believed neutral state. Our study establishes oxygen vacancies as the dominant donor-like native point defects in ZnO and well explains the unintentional n-type conductivity, a mysterious problem in the last fifty-year study of ZnO. The methodology we have developed can be used to study oxygen-related point defects and their energetics in other technologically important metal oxides such as TiO$_2$ and In$_2$O$_3$, and thus is of general interest in oxide and semiconductor physics community.

## Methods

Isotopically modulated samples were grown by radio-frequency plasma assisted molecular beam epitaxy (rf-MBE) on c-oriented sapphire substrate with several-nanometer-thick nitrided structure on the surface, and more nitridation pretreatment details can be found elsewhere[35]. The base pressure was in the range of $10^{-9}$ mbar. Two sorts of oxygen sources were used in the synthesis, one in its natural isotopic ratio (labeled as $^n$O) and another one artificially enriched with $^{18}$O isotope (> 99.7%). A Zn$^n$O buffer layer was firstly deposited at 723 K, providing a good epitaxial template. Further, a Zn$^n$O epitaxial layer and an $^{18}$O-enriched Zn$^{18}$O layer were subsequently grown at 873 K. To realize the study of diffusion profiles as a function of $\mu$, the Zn flux was varied for different samples, while the radio frequency power of the plasma gun and the flux of oxygen gas were kept constant – 300 W and 2.0 standard cubic centimeter (sccm), respectively. Specifically, the temperature of Zn K-cell for Z1 – Z5 is 605 K, 591 K, 585 K, 579 K and 569 K, respectively. Thus, a series of isotopic heterostructures consisting of top Zn$^{18}$O and bottom Zn$^n$O were synthesized. It should be noted that the $^{18}$O source used in the top Zn$^{18}$O layer was mixed with 1.0 sccm artificially enriched $^{18}$O (> 99.7%) and 1.0 sccm $^n$O, and the chemical potentials of Zn and O were respectively kept the same for the two different epilayers in each sample. Reflection high-energy electron diffraction (RHEED) was utilized *in-situ* to monitor the whole epitaxial growth processes (Supplementary information Fig. S8).

Prior to the diffusion anneal, the surfaces of two identical samples were brought together in a diffusion couple to minimize the evaporation. Annealing experiments were performed in a temperature range of 1023 K – 1148 K for 30 min in vacuum. In order to ensure a fast ramping up to the desired diffusion temperature, the removable furnace was preheated and then moved to the sample place to perform the annealing process. The annealing temperatures were controlled within $\pm 2$ K. After annealing, the furnace was moved away immediately. Note that the annealing temperature range is 1023 K – 1173 K for the two more batches of samples with modulated Fermi level – Z1 to Z1-2 and Z2 to Z2-2.

Concentration depth profiles were measured by secondary ion mass spectroscopy (SIMS) using a Hiden MAXIM Analyser. For SIMS analysis 5 keV Ar$^+$ ions with a current 220 nA was used as a primary beam raster over an area of $200 \times 200$ μm$^2$. The signal-to-concentration calibration was performed using standard $^n$O and $^{18}$O samples as reference. The conversion of SIMS sputtering time to depth profiles was performed by measuring the crater depth using a Dektak 8 profilometer and assuming a constant erosion rate. Temperature-dependent Hall measurements were performed by HL5500PC Hall Effect Measurement System.

Note that the error bars in Fig. 3 - 5 can be explained by the inaccuracy originated from the uncertainty in SIMS measurements. The fact that SIMS sputtering rate is not being constant will lead to a maximum error of 20% in *D*. The analysis of $D_O$ and $\Delta H_a$ indicates that the values are within the acceptable limit and supports our above discussions (Supplementary Information S9).

## Acknowledgements

This work was supported by the Ministry of Science and Technology of China (Grant nos 2011CB302002, 2011CB302006), the National Science Foundation of China (Grant nos 11174348, 51272280, 11274366, 61204067, 61306011, 11427903), the Chinese Academy of Sciences, and the Research Council of Norway (Grant no 221668).

## Author Contributions

Z.M., Q.X. and X.D. conceived and guided the study. L.L. and Z.M. designed the research. L.L. and A.H. conducted the growth and characterization. L.L., Z.M., X.D. A.A. A.K. and Q.X. did the analysis. L.L. wrote the paper. All authors discussed the results and commented to the manuscript.

## Additional Information

Supplementary information is available from Online or from the author.

## Competing financial interests

The authors declare no competing financial interests.